\documentclass[conference,a4paper]{IEEEtran}
\IEEEoverridecommandlockouts
\usepackage{stfloats,color}
\usepackage{amsfonts}
\usepackage[cmex10]{amsmath}
\usepackage{graphicx}
\usepackage{setspace}
\usepackage{cite}
\usepackage{array}
\usepackage{algorithmic}
\usepackage{mdwmath}
\usepackage{mdwtab}
\usepackage[center]{caption}
\usepackage[font=footnotesize]{subfig}
\usepackage{fixltx2e}
\usepackage{url}
\usepackage{times}
\usepackage{epsfig}
\usepackage{latexsym}
\usepackage{epstopdf}
\usepackage{verbatim}
\usepackage{units}
\usepackage{amsthm}
\usepackage{mdwlist}

\newtheorem*{theorem:outage}{Definition}

\begin{document}
\title{On the Role of Vehicular Mobility in \\ Cooperative Content Caching}

\author{\IEEEauthorblockN{Osama Attia, Tamer ElBatt}
\IEEEauthorblockA{
Wireless Intelligent Networks Center (WINC)\\
Nile University, Cairo, Egypt 12677\\
\{\tt osama.gamal@nileu.edu.eg, telbatt@ieee.org\}}
\thanks{This work was funded in part by a research grant from General Motors Company.}
}

\maketitle

\begin{abstract}
\boldmath
In this paper, we analyze the performance of cooperative content caching in vehicular ad hoc networks (VANETs). In particular, we characterize, using analysis and simulations, the behavior of the probability of outage (i.e. not finding a requested data chunk at a neighbor) under freeway vehicular mobility. First, we introduce a formal definition for the probability of outage in the context of cooperative content caching. Second, we characterize, analytically, the outage probability under vehicular and random mobility scenarios. Next, we verify the analytical results using simulations and compare the performance under a number of plausible mobility scenarios. This provides key insights into the problem and the involved trade-offs and enable us to assess the potential opportunity offered by the, somewhat structured, vehicular mobility that can be exploited by cooperative content caching schemes. The presented numerical results exhibit complete agreement between the analytical and simulation studies. Finally, we observe that vehicular mobility creates opportunities for enhanced outage performance under practically relevant scenarios.\\
\end{abstract}


\IEEEpeerreviewmaketitle

\section{Introduction}
\IEEEPARstart{V}\small{e}hicular Ad hoc Networks (VANET) is a promising networking paradigm that received increasing interest from the academic and industry research communities over the last decade. This is primarily attributed to its key role in improving the driving experience and saving lives on the roads. VANETs are envisioned to support vehicle-to-vehicle (V2V) and vehicle-to-infrastructure (V2I) communications, depending on the automotive applications and services of interest. They are projected to leverage multiple wireless access technologies with diverse data rate, radio range and mobility support capabilities, e.g., IEEE 802.11g/a/p, IEEE 1609, and IEEE 802.16e.

Content caching schemes have been first introduced to the Internet, particularly for web caching, e.g., \cite{wang1999survey,Barish:2000www}, and more recently to its wireless and mobile extensions \cite{Niesen:2009ws} in order to relieve the load on the network and, hence, improve its throughput and delay performance. There are two major paradigms for caching, namely non-cooperative caching (or simply caching) \cite{Cao:2002pro} and cooperative caching \cite{Jing:2010coo}. Under the former paradigm, nodes take independent, uncoordinated decisions as to which currently consumed data or routing paths should be cached and for how long. Under the more recent cooperative caching paradigm, nodes may share cached data and jointly take decisions as to what information may be cached and for how long. Unlike non-cooperative caching, cooperative caching has been shown to exploit the wisdom of the crowd creating content diversity and utilizing the nodes' storage in a more efficient manner \cite{sailhan:2003cooperative}. This paper lies at the intersection of the two aforementioned research areas, largely studied in isolation, namely vehicular networking and cooperative content caching. In particular, we explore the benefits, and potential trade-offs, the structured vehicular mobility might bring to the performance of cooperative content caching.

Recent work has focused on modeling vehicular mobility, to better understand its effect in a variety of settings \cite{Saha:2004mmv,Fiore:2007wq,Durrani:vx,Sommer:2008wz}. In particular, developing models for use in a variety of network simulation tools, e.g., {\it ns-2}, Qualnet \cite{Hadi:2010uz,Mahajan:2006tx,Harri:2006:VGR}. In \cite{Fiore:2007wq}, vehicular mobility models in both highway and urban scenarios were introduced and tested using network simulation studies showing their impact on the performance of routing protocols. However, the effect of vehicular mobility on cooperative content caching was not explored.

On the other hand, cooperative caching in wireless ad hoc networks has been studied in \cite{Yin:2004we,Fiore:2009vf}. \cite{Yin:2004we} proposes three schemes for cooperative caching with the objective of reducing the query delay, yet, without focus on various node mobility patterns or their potential impact on cooperative content caching performance. In \cite{Fiore:2009vf}, the authors also consider cooperative content caching and introduce a new metric, namely the {\it presence index}, deciding for how long should a data chunk be cached (cache drop time) aiming for an enhanced queue resolution probability. In \cite{Fiore:2009vf}, the authors consider two mobility scenarios, namely pedestrian mobility in a mall and vehicular mobility on a city road grid, yet, the structured nature of vehicular mobility, particularly on freeways, has not been exploited to improve the query resolution probability.

Cooperative content caching yields outage probability gains, referred to as cooperation diversity gains, attributed to the multiple copies of a data chunk available at neighboring nodes. In vehicular scenarios, we argue that the structured and predictable mobility patterns of vehicles on roads bring about another form of outage probability gains, referred to as mobility gains. In this paper, we focus on the gains attributed to mobility while the cooperation diversity lies out of the scope of this paper and is a subject matter of  ongoing research.

Our main contribution in this paper is two-fold. First, we introduce and formally define the probability of outage as a performance metric for cooperative content caching. Second, we conduct outage analysis under vehicular and random mobility scenarios. Our model and the accompanying analysis help us identify scenarios under which vehicular mobility creates an ample opportunity for performance improvement over random mobility regimes dominant in the mobile ad hoc networks (MANET) literature.

The rest of this paper is organized as follows. In Section \ref{sec:sys_model} we present the system model and underlying assumptions. Section \ref{sec:outage} introduces the probability of outage performance metric and presents analysis comparing the outage behavior under random and vehicular mobility regimes. Performance results are presented and discussed in Section \ref{sec:results}. Finally, in Section \ref{sec:conclusion}, conclusions are drawn and potential directions for future work are pointed out.

\section{System Model \label{sec:sys_model}}
In this section, we introduce the system model and assumptions underlying our analytical and simulation studies. First, we give the assumptions underlying the cooperative content caching system. Afterwards, we introduce the vehicular and random mobility models considered in this paper.

The following assumptions are made about the system under investigation:
\begin{itemize*}
\item We assume that vehicular users are interested in information items where each item consists of $C$ chunks.
\item We analyze the performance of cooperative caching on a system of two nodes. The rationale behind this is two-fold: (i) The system of two nodes fully captures the impact of mobility on cooperation which is the prime focus of this paper and (ii) It allows us to conduct a thorough analysis and capture fundamental insights, due to the mathematical tractability of the model.
\item Assume that the two nodes start at time $t=0$ with empty caches.
\item A node requests a chunk that is not stored in its cache. Chunk requests arrive at node $i$ according to a Poisson process with rate $\lambda_i$ = $\lambda$.
\item Assume the request overlap ratio between two neighbors (modeling their common interest) is given by the parameter $\gamma$, where $0 \leq \gamma \leq 1$. For the two node model studied in this paper, we assume $\gamma$ is fixed throughout.
\item Assume that each node uses a fixed transmit power which translates to a circular range of radius $r$.
\item If the requesting node gets a query resolved, it caches a copy of the chunk for an arbitrarily long time. This is justified by our prime focus on capturing the sole impact of vehicular mobility on the cooperative caching performance. Incorporating the effect of finite caching times lies out of the scope of this paper and is a subject of future research.
\end{itemize*}

In order to study the impact of nodes' mobility on the performance of cooperative content caching, we discuss next two generic mobility models that capture the crux of vehicular mobility on a straight segment of a freeway and multiple variations of random mobility. This is of paramount importance to gain valuable insights about the benefits, and potential trade-offs, of structured (as opposed to random) mobility. These insights give rise to fundamental limits and principles, which are the subject matter of future research, that govern the design of future cooperative content caching schemes to best exploit the structured nature of vehicular mobility.

\subsection{Random Mobility Model}
\begin{figure}[!t]
\centering
\includegraphics[width=0.35\textwidth]{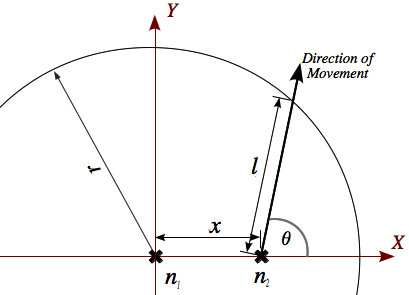}
\setlength{\abovecaptionskip}{0pt}
\setlength{\belowcaptionskip}{-10pt}
\caption{Illustration of the random mobility model}
\label{Flo:model}
\end{figure}

Under random mobility, we assume two nodes, a stationary node denoted $n_1$ at the origin and a mobile node denoted $n_2$ located at $(x,0)$ at time $0$, as shown in Fig. \ref{Flo:model}. This is attributed to our focus on the relative motion which governs the node's neighborhood relationship. As pointed out earlier, $n_2$ is assumed to be initially located at $(x,0)$ and is moving with speed $v$ along a direction of angle $\theta$ from the +ve X-axis. $x$, $\theta$ and $v$ are assumed to be independent random variables that are uniformly distributed over the ranges $[-r,r]$, $[\theta_{min}, \theta_{max}]$ and $[v_{min}, v_{max}]$, respectively.  We study different scenarios that reflect a variety of random mobility patterns ranging from the somewhat theoretical random mobility with $\theta$ covering the entire two-dimensional space (i.e. $\theta \sim \mathrm{U}[0, 2\pi]$) to random mobility with limited $\theta$ range.

\subsection{Freeway Vehicular Mobility Model}
Due to the structured vehicular motion, the freeway vehicular mobility is considered as a special case of the random mobility model discussed above. We consider a straight freeway segment along the Y-axis accommodating two nodes, where $n_1$ is located  at the origin and $n_2$ is the mobile node located at $(x,0)$ at time $0$ and moving along the Y-axis with speed $v$. Obviously, $\theta$ in the straight freeway vehicular mobility regime, is assumed deterministic and equals to $\nicefrac{\pi}{2}$ (or $- \nicefrac{\pi}{2}$). On the other hand, $x$, and $v$ are independent uniformly distributed random variables over $[-r,r]$, and $[v_{min}, v_{max}]$, respectively, as in the random mobility model.

In the next section, we conduct the outage performance analysis for cooperative content caching under, both, random and vehicular mobility regimes.

\section{Outage Performance Analysis \label{sec:outage}}
\subsection{Probability of Outage}
In this section, we introduce the metric adopted to quantify and compare the performance of cooperative content caching under a variety of mobility regimes, namely the outage probability.

An outage event occurs when a node cannot retrieve a requested chunk from a single-hop neighbor within a period of length $\tau$. Conceptually, it can be generalized to two-, three- or {\it k-hop} neighbors, however, we limit our attention in this first look at the problem to cooperation among single-hop neighbors.

\begin{theorem:outage}
We define the probability of outage as the probability of not finding a data chunk at a single-hop neighbor within time period $(t, t+\tau)$.
\end{theorem:outage}

Formally, the outage probability at node $n_1$, denoted $P_o^{n_1}$, can be defined as the complement of the probability of node $n_1$ finding a chunk, denoted $P_f^{n_1}$, that is,
\begin{align*}
P_o^{n_1} &= 1 - P_f^{n_1}
\end{align*}

\noindent where the event of $n_1$ finding a requested data chunk at the neighbor $n_2$ within time period $\tau$ occurs when $n_2$ requests at least a chunk within the period $\tau$, there is an interest overlap, among the requested chunks, between the two nodes $\gamma$ and the two nodes are within the communication range, that is,
\begin{align}
P_f^{n_1} = & \gamma ~ P_{neigh} ~ P_r^{n_2} \label{eqn:finding}
\end{align}

\noindent where $P_r^{n_2}$ is defined as the probability of $n_2$ requesting at least a data chunk within the period $\tau$. Based on the Poisson arrival of requests, it is given by,
\begin{align*}
P_r^{n_2} = 1 - e^{-\lambda \tau}
\end{align*}

Furthermore, $P_{neigh}$ is the probability of nodes $n_1, n_2$ are within the communication range after time $\tau$.

It is evident from \eqref{eqn:finding} above, that the mobility effect prevails through the probability of the two nodes being within reach after $\tau$, $P_{neigh}$. Thus, in the next subsection, we shed some light on the behavior of $P_{neigh}$ under different mobility scenarios in an attempt to better understand its impact on the performance of cooperative content caching.

\subsection{Analysis}
First, we define relevant variables and notation for the general setting and then move to analyze the two mobility regimes of interest. We assume, without loss of generality, that all distances are normalized by $r$. Thus, the initial distance between the two nodes, denoted $x$, is uniformly distributed over $[-1, 1]$. For notation convenience, we introduce a new random variable, $u$, as a function of the velocity random variable $v$ such that $u = v\tau$. Let $l$ be the distance that $n_2$ needs to cover to reach the maximum transmission range of $n_1$. Hence, $n_2$ will stay within the radio range of $n_1$ after time $\tau$ {\it iff} if $v\tau$ is less than or equal to distance $l$, that is $u \le l$. For an arbitrary $\theta$, $l$ is given by,
\begin{align}
l = \sqrt{1-x^2 \sin^2{\theta}} - x \cos{\theta}
\end{align}

\subsubsection{Random Mobility}
It is evident from the discussion above that $P_{neigh} = P(u \le l)$, that is,
\begin{align}
P_{neigh} &= P(u \leq l) \notag \\ &= P(u \leq \sqrt{1-x^2 \sin^2{\theta}} - x \cos{\theta}) \label{eqn:prob_rand_mob}
\end{align}

\begin{figure}[!t]
\centering
\includegraphics[width=0.36\textwidth]{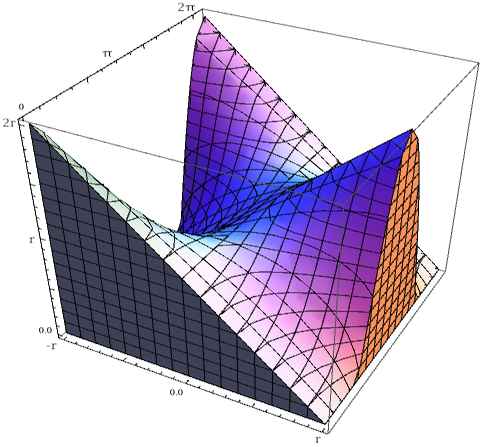}
\setlength{\abovecaptionskip}{0pt}
\setlength{\belowcaptionskip}{-10pt}
\caption{The region $D_r$ for the event in \eqref{eqn:prob_rand_mob} under the random mobility model}
\label{Flo:region3d}
\end{figure}

The event in \eqref{eqn:prob_rand_mob} depends on three independent random variables, namely $x$, $u$, $\theta$ and its probability is determined as follows,
\begin{align}
P_{neigh} = \iiint_{x,u,\theta \in D_r}{f(x,u,\theta)}\mathrm{d}x\mathrm{d}u\mathrm{d}\theta \label{eqn:rand_mob}
\end{align}
\noindent where $f(x,u,\theta)$ is the joint probability density function (PDF) of $x$, $u$, and $\theta$ which boils down to the product of their marginal PDFs due to independence. $D_r$ is the region characterizing the random event in \eqref{eqn:prob_rand_mob} which lies underneath the curve shown in Fig. \ref{Flo:region3d}. Unfortunately, the integration in \eqref{eqn:rand_mob} turns out to be complex and cannot be solved in closed form. Therefore, the probability of interest, $P_{neigh}$, is quantified numerically as demonstrated in Section \ref{sec:results}.
\begin{figure}[!t]
\centering
\includegraphics[width=0.41\textwidth]{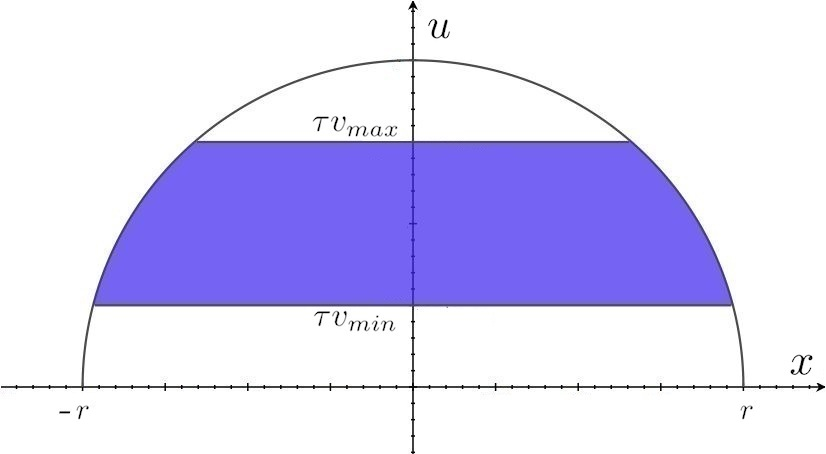}
\setlength{\abovecaptionskip}{5pt}
\setlength{\belowcaptionskip}{-20pt}
\caption{The shaded region $D_v$ for the event in \eqref{eqn:veh_mob} under the vehicular mobility model}
\label{Flo:region}
\end{figure}

\subsubsection{Vehicular Mobility}
The outage analysis in case of vehicular mobility is relatively simple and goes along the same lines of the random mobility case. In fact, its a special case because the angle of movement, $\theta$, is deterministic due to the straight mobility along the Y-axis aligned freeway. This, in turn, yields $l = \sqrt{1-x^2}$, in the straight freeway mobility scenario since $\theta$ is deterministic and is given by $\theta = \nicefrac{\pi}{2}$. Hence:
\begin{align}
P_{neigh} = P(\tau v_{min} \leq u \leq \min(\tau v_{max}, \sqrt{1-x^2})) \label{eqn:prob_veh_mob}
\end{align}
which is a function of two independent random variables, namely $x$, $y$ and is given by,
\begin{align}
P_{neigh} &= \iint_{x,u\in D_v}{f(x,u)}\mathrm{d}x\mathrm{d}u \label{eqn:veh_mob}
\end{align}
\noindent where $D_v$ is the region over which $x$ and $u$ satisfy the inequality $\tau v_{min} \leq u \leq \min(\tau v_{max}, \sqrt{1-x^2})$ as in Fig. \ref{Flo:region}.\\*

In the next section, we present results comparing three variants of the random mobility model to the straight freeway vehicular mobility model.

\section{Performance Results \label{sec:results}}
In this section, we present analysis and simulation results illustrating the effect of mobility on the outage performance of cooperative content caching. We develop Matlab simulations to verify the analytical results. Analytical and simulation results are generated using the following system parameters: $\gamma = 0.7$, $\lambda = 3~requests/sec$, $r = 150~m$, $v_{min} = 5~km/hr$, and $v_max = 50~km/hr$. Next, we present comparative results under three mobility scenarios.

\subsection{Random Mobility\label{subsec:rand_mob_res}}
First, we analyze the random mobility model described in Section \ref{sec:sys_model} with $\theta \sim \mathrm{U}[0, 2\pi]$ along with the parameters given above. According to \eqref{eqn:rand_mob}, the probability of being in reach in this case, $P_{neigh}$, would be given by,
\begin{align}
P_{neigh} &= \int_{0}^{2\pi}{\int_{u_{min}}^{u_{max}}{\frac{\sqrt{1-u^2\sin^2{\theta}}-u\cos{\theta}}{4\pi(u_{max}-u_{min})}~\mathrm{d}u}~\mathrm{d}\theta}
\end{align}
On the other hand, $P_{neigh}$ under vehicular mobility follows from \eqref{eqn:veh_mob} and is given by,
\begin{align}
P_{neigh} &= \int_{u_{min}}^{u_{max}}{\frac{\sqrt{1-u^2}}{(u_{max}-u_{min})}}~\mathrm{d}u
\end{align}

\begin{figure}[!t]
\centering
\subfloat[Probability of being in reach under Random and Vehicular Mobility]{
\includegraphics[width=0.5\textwidth]{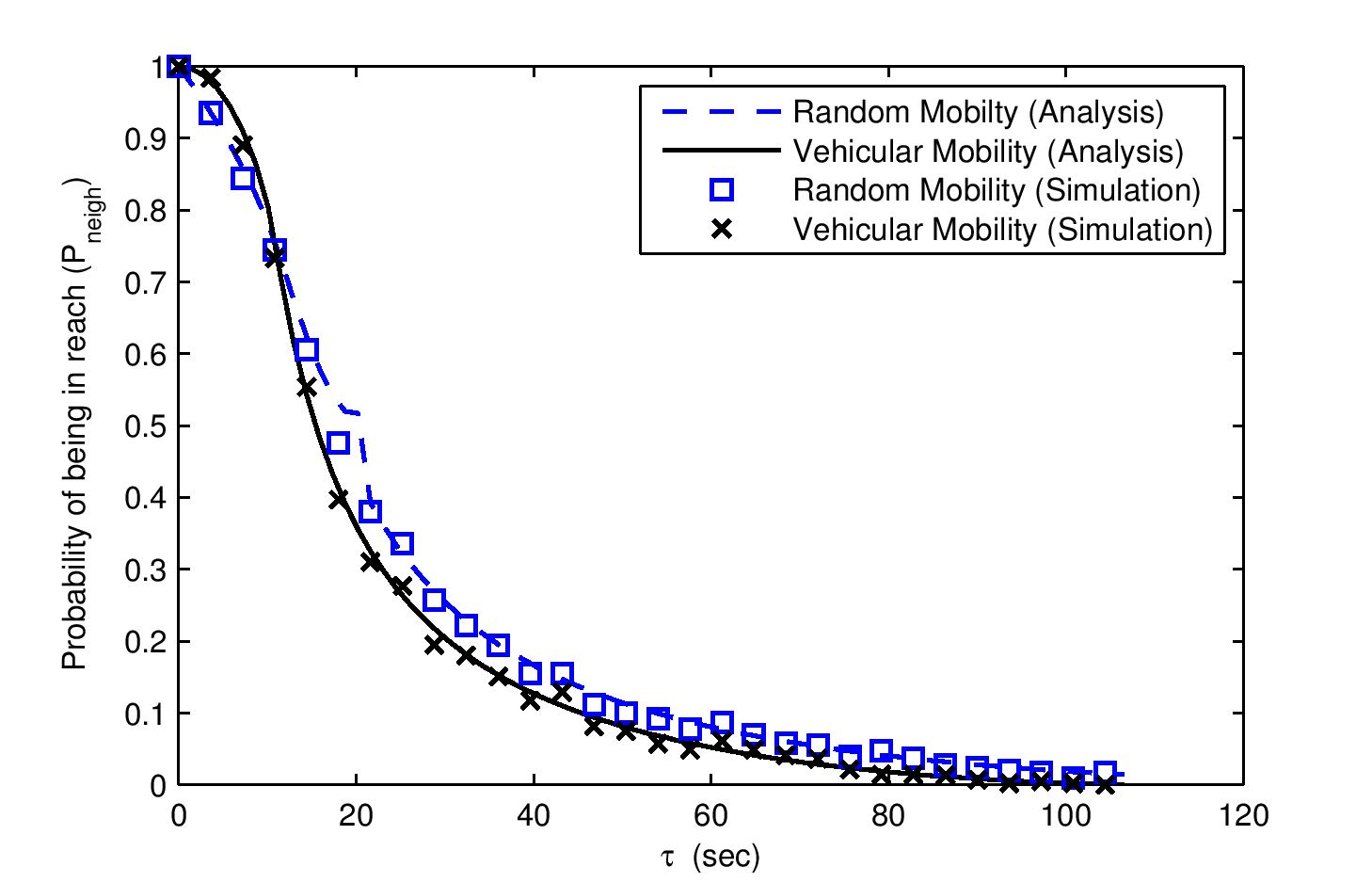}
\label{Fig:comp_rand_being}
\setlength{\abovecaptionskip}{-10pt}
\setlength{\belowcaptionskip}{-10pt}
}\\
\subfloat[Outage Probability under Random and Vehicular Mobility]{
\includegraphics[width=0.5\textwidth]{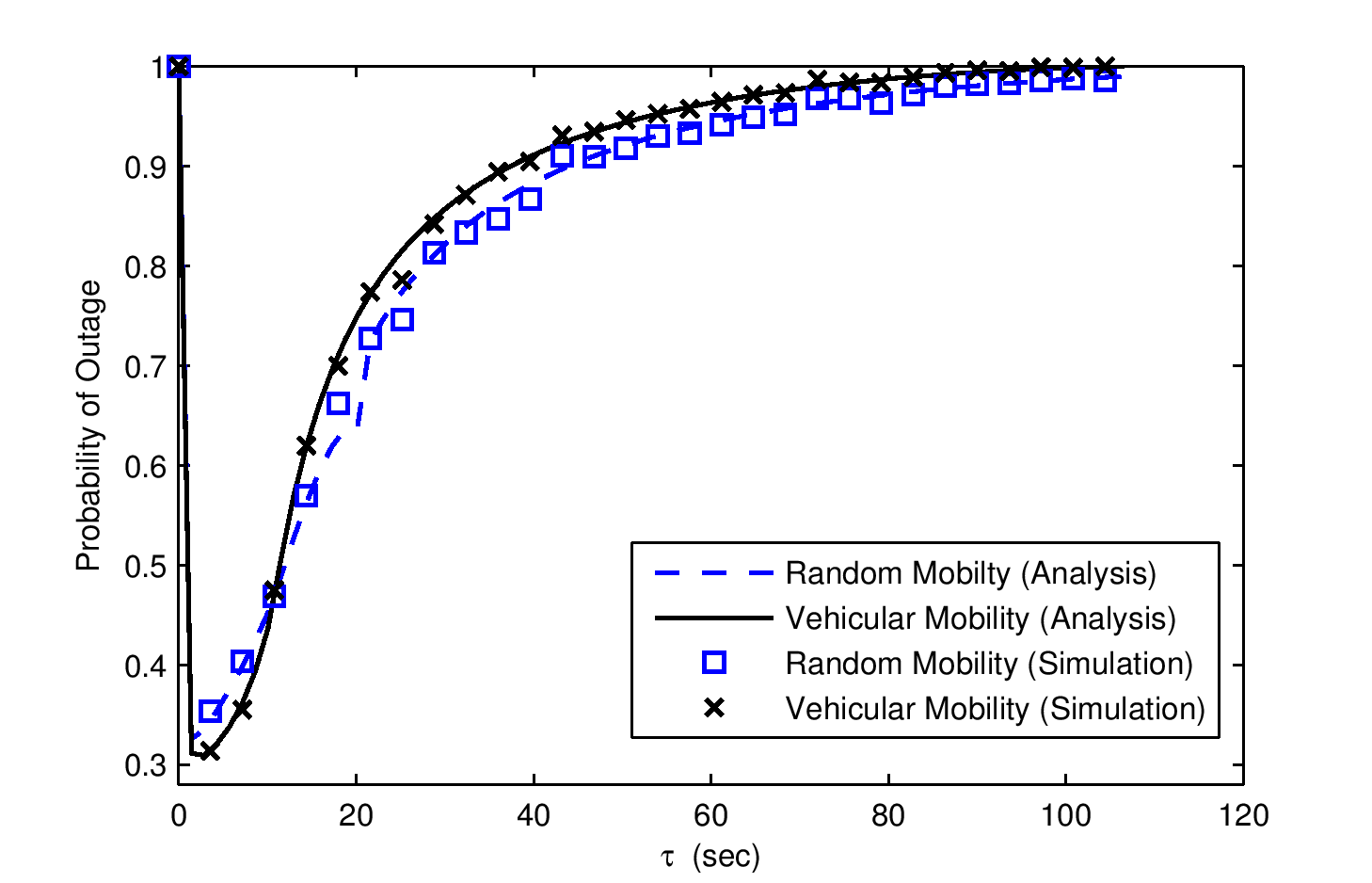}
\label{Fig:comp_rand}
\setlength{\abovecaptionskip}{-10pt}
\setlength{\belowcaptionskip}{-10pt}
}
\caption{Random mobility, where $x\sim\mathrm{U}[-r,r]$, $\theta\sim\mathrm{U}[0,2\pi]$, $v\sim[v_{min}, v_{max}]$}
\setlength{\abovecaptionskip}{-10pt}
\setlength{\belowcaptionskip}{-15pt}
\end{figure}

\begin{figure}[!t]
\centering
\advance\leftskip4pt
\includegraphics[width=0.5\textwidth]{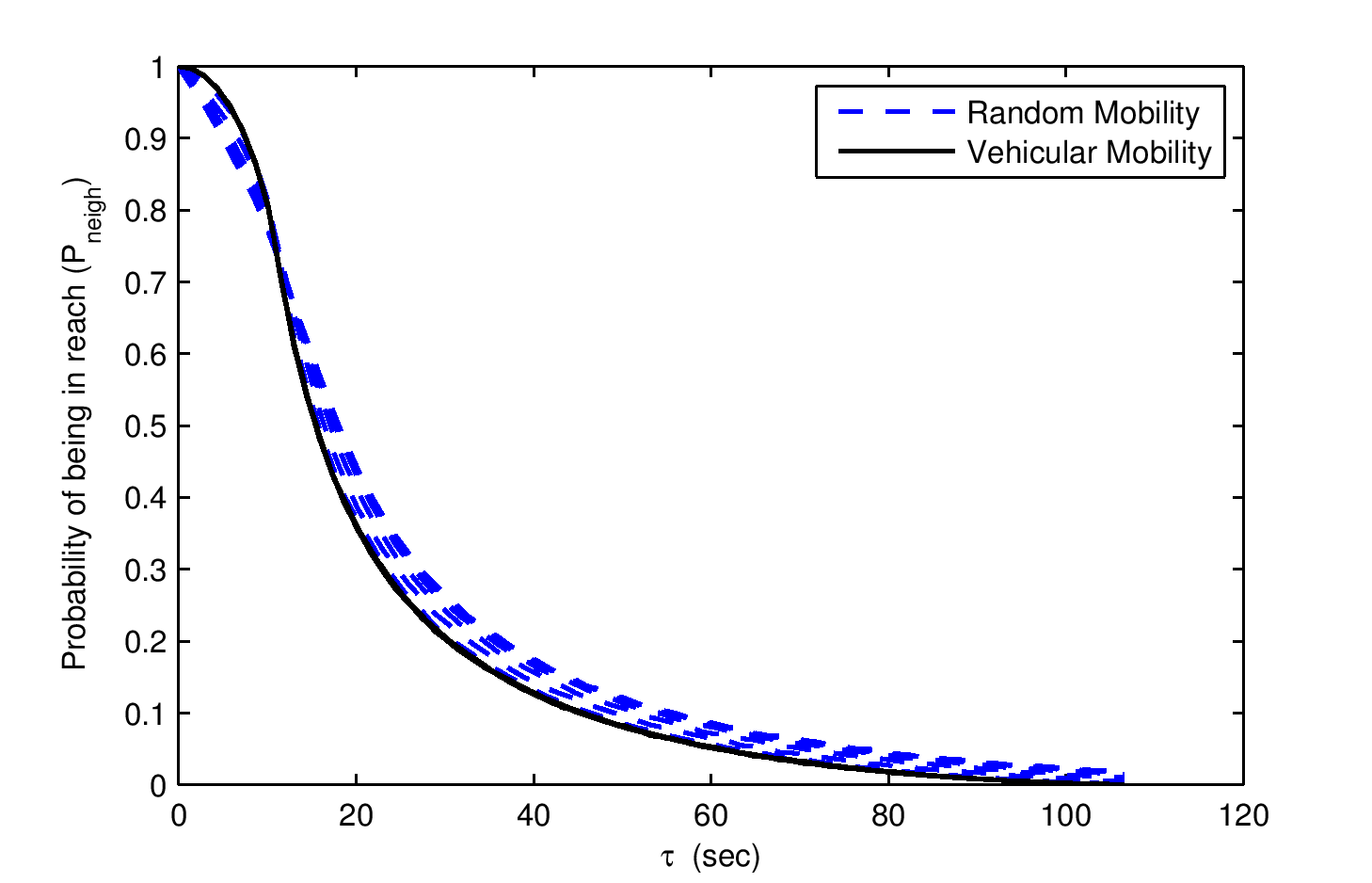}
\setlength{\abovecaptionskip}{-10pt}
\setlength{\belowcaptionskip}{-15pt}
\caption{Illustrating the effect of changing the range of $\theta$ on \\ Random Mobility}
\label{Fig:comp_rand_around_y}
\end{figure}

Due to the complexity of the integration in the random mobility case, being not amenable to a closed form solution, we embraced numerical techniques yielding the results of $P_{neigh}$ shown in Fig. \ref{Fig:comp_rand_being}. Substituting in \eqref{eqn:finding} results Fig. \ref{Fig:comp_rand}, where the vehicular mobility has lower probability of outage than the random mobility for ranges of $\tau$ where the outage probability is of practical interest (e.g., $P_o \le 0.1$). For small $\tau$, both curves start with high outage probabilities due to the low probability of $n_2$ requesting at least one chunk within a short period $\tau$. Afterwards, the probability of outage becomes considerably low due to the contribution of high probability of being in reach. Finally, it increases again due to the decreasing probability of being in reach between $n_1$ and $n_2$, which starts to dominate performance, as the period $\tau$ gets longer. In addition, Fig. \ref{Fig:comp_rand} shows Matlab simulation results exhibiting complete agreement with the analytical results. Thus, we rely on simulation results for the rest of the paper to analyze more general mobility scenarios for which characterizing the integration region, $D_r$, in \eqref{eqn:rand_mob} becomes more complex.
The above trends and relative performance witnessed in Fig. \ref{Fig:comp_rand} (esp. the crossover) motivated us to take a closer look at variants of the studied random mobility, in comparison to vehicular mobility, in an attempt to better understand the trade-offs involved. Hence, in Fig. \ref{Fig:comp_rand_around_y}, we show using Matlab simulations that shrinking the range of the random variable $\theta$ (direction of motion) symmetrically around the Y-axis eventually converges to the vehicular mobility around the Y-axis. Thus, our key observation here is that the phenomena of "crossover" between the mobility and vehicular curves persists for all ranges of $\theta$. However, for small values of $\tau$, where $P_o$ is of practical interest, vehicular mobility always outperforms random mobility.

\subsection{Limited Random Mobility}
\begin{figure}[!t]
\centering
\includegraphics[width=0.5\textwidth]{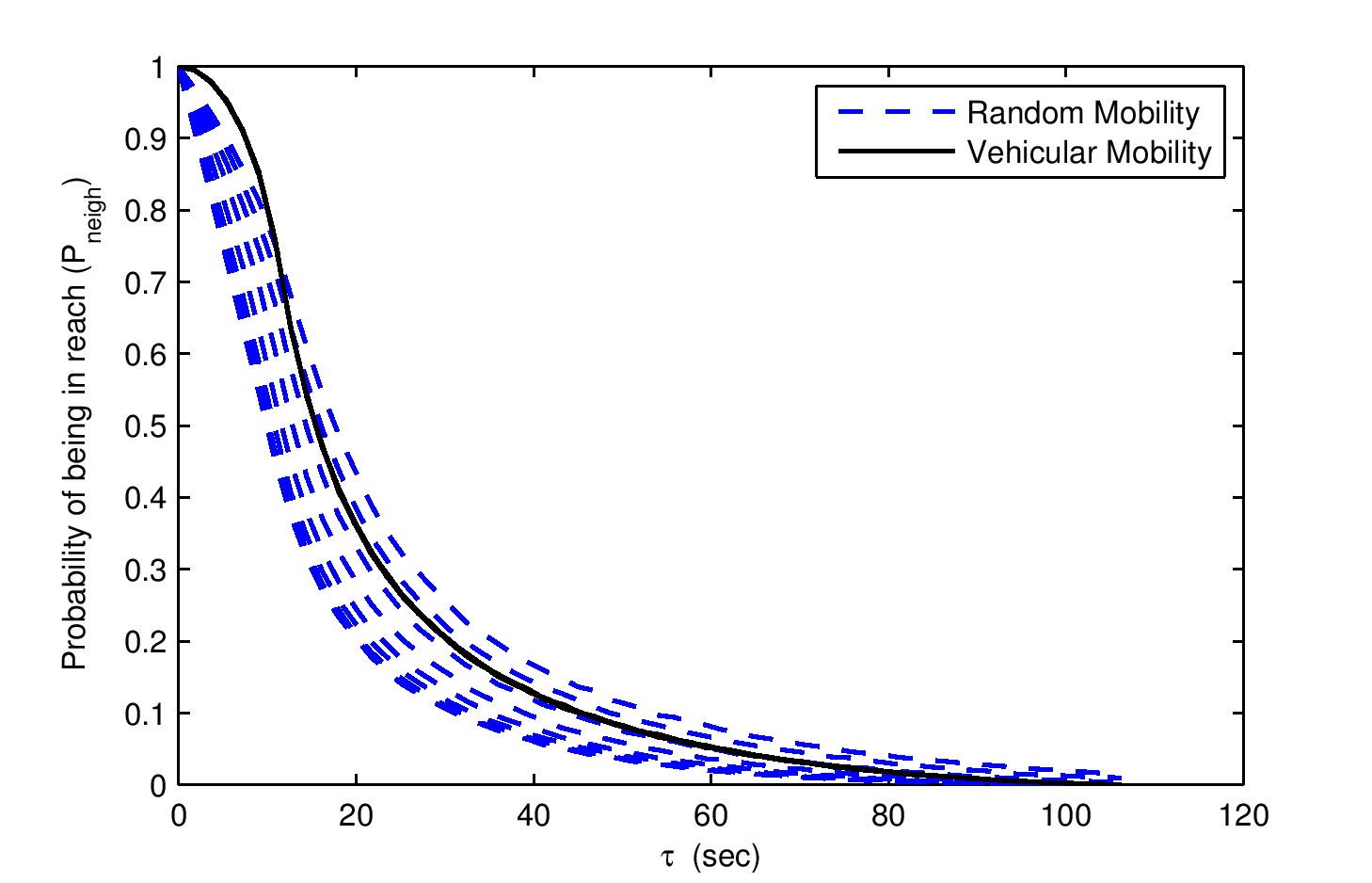}
\setlength{\abovecaptionskip}{-10pt}
\setlength{\belowcaptionskip}{-10pt}
\caption{Illustrating the effect of changing the range \\ of $\theta$ on limited Random Mobility}
\label{Fig:lim_rand_around_x}
\end{figure}

Another interesting scenario is the one where we constrained the initial position, $x$, in both cases to be uniformly distributed over $[0,r]$ cancelling the symmetry around the Y-axis. This provides further insights into the contribution of this symmetry to the crossover behavior. Fig. \ref{Fig:lim_rand_around_x} shows the effect of shrinking the range of $\theta$ from $[0, 2\pi]$ to be exactly along the positive X-axis, i.e. orthogonal to the vehicular mobility. This, in turn, gives rise to two key observations. First, vehicular mobility outperforms random mobility by $77.9\%$, on the average. Second, the crossover point got shifted to the right (due to the symmetry around the X-axis) providing a wider range of $\tau$ over which vehicular mobility yields gains.

\subsection{Realistic Vehicular Mobility}
\begin{figure}[!t]
\centering
\includegraphics[width=0.5\textwidth]{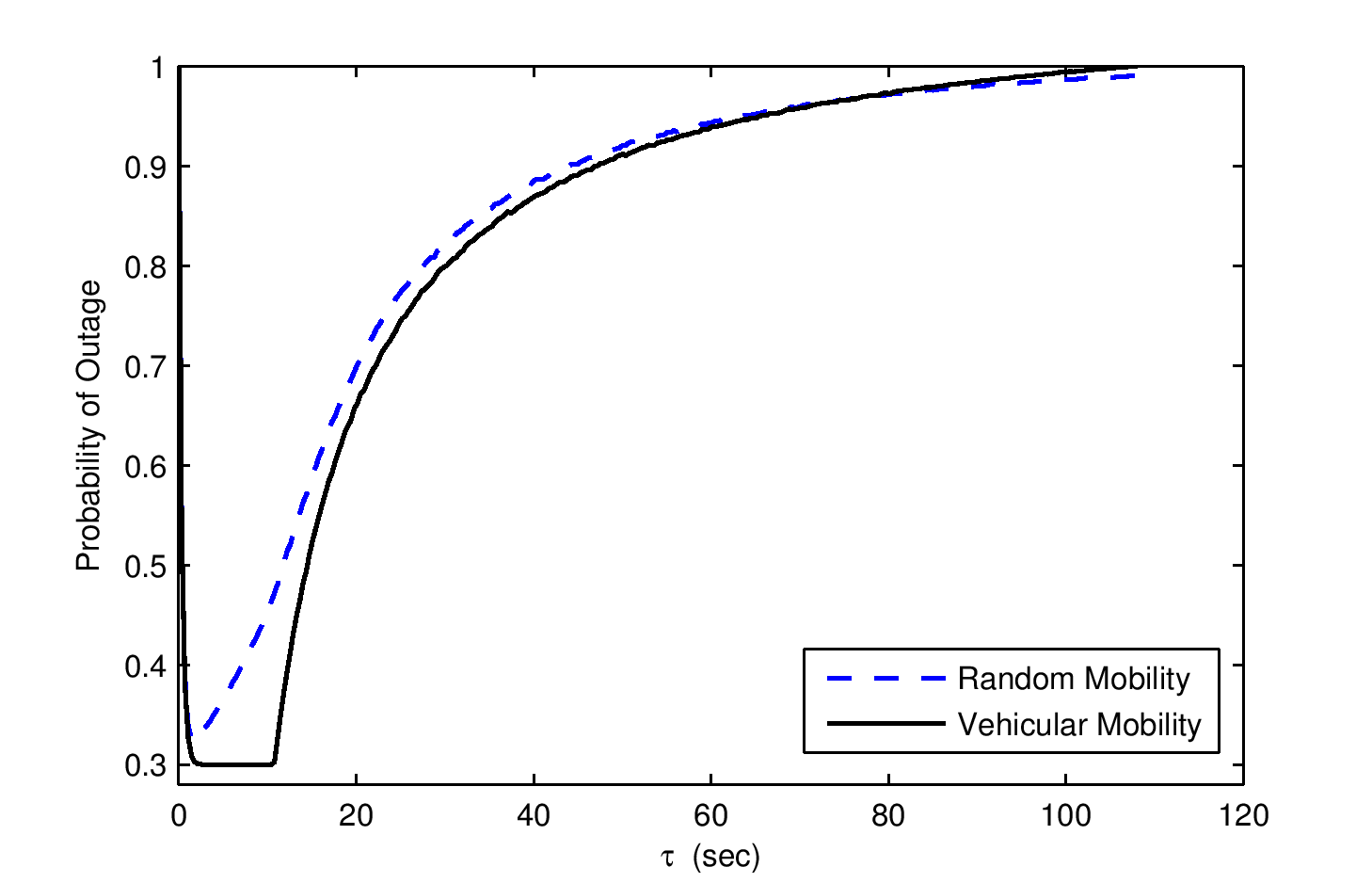}
\setlength{\abovecaptionskip}{-10pt}
\setlength{\belowcaptionskip}{-10pt}
\caption{Towards Realistic Freeway Mobility}
\label{Fig:realistic}
\end{figure}

This scenario constitutes an attempt towards approaching a realistic mobility scenario where the initial vehicle separation, $x$, in the vehicular mobility scenario is uniformly distributed over the width of the freeway. On the other hand, random mobility follows the model described in Section \ref{subsec:rand_mob_res}. Accordingly, Fig. \ref{Fig:realistic}, once more, confirms the superiority of  vehicular mobility with respect to outage performance, e	specially in the practical range of interest (i.e. $\tau \le 10 sec$). Furthermore, the scenario leads to a maximum improvement of $32\%$ and the crossover point shifts right due to the contribution of the limited freeway width (assuming a 5-lane freeway with 4 meters lane width).
%

\section{Conclusion \label{sec:conclusion}}
We analyzed the performance of cooperative content caching in vehicular ad hoc networks. More specifically, we characterized the behavior of the outage probability (i.e. not finding a requested data chunk) under vehicular and random mobility regimes. First, we introduced a formal definition for the probability of outage in the context of cooperative content caching. Second, we characterized, analytically, the outage probability under vehicular and random mobility. We verified the analytical results using simulation studies which exhibit complete agreement. The presented results confirm the opportunity created by the structured vehicular mobility which would inspire future cooperative caching schemes. The numerical results demonstrate up to $32\%$ improvement in the outage performance (and $16\%$ on the average) for the studied plausible scenarios where the probability of outage is below $0.5$. This work can be extended along the following research directions: (i) incorporate more general and realistic vehicular mobility patterns, e.g., city and intersection scenarios, (ii) explore another dimension of mobility gains, namely mobility diversity contributed by the new content brought by vehicles passing by (this is captured by the {\it average} content overlap factor $\gamma$ which becomes time-varying in this setting), (iii) quantify the cooperation diversity gain that is above and beyond the mobility gains explored here and (iv) develop novel cooperative caching schemes that capture the structured nature of vehicular mobility.

\linespread{1.1}



\end{document}